\documentclass[preprint,12pt, %showpacs, 
aps, %tightenlines,
amsmath,amssymb,nofootinbib,superscriptaddress,hyperref
]{revtex4-1} 
\usepackage[utf8]{inputenc}
\usepackage{epsfig} 
\usepackage{wasysym} 
\usepackage{mathrsfs} 
\usepackage{graphicx} 
\usepackage{epstopdf}
\usepackage{amsfonts} 
\usepackage{amsbsy} 
\usepackage{amscd} 
\usepackage{pstricks} 
\usepackage{multirow}
\usepackage{slashed} 
\usepackage{tikz}
\usetikzlibrary{decorations.pathmorphing,decorations.markings}

\newcommand{\beq}{\begin{equation}}
\newcommand{\eeq}{\end{equation}}
\newcommand{\bea}{\begin{eqnarray}}
\newcommand{\eea}{\end{eqnarray}}
\newcommand{\beas}{\begin{eqnarray*}}
\newcommand{\eeas}{\end{eqnarray*}}
\newcommand{\bi}{\begin{itemize}}
\newcommand{\ei}{\end{itemize}}

\def\tev{\,{\ifmmode\mathrm {TeV}\else TeV\fi}}
\def\gev{\,{\ifmmode\mathrm {GeV}\else GeV\fi}}
\def\to{\rightarrow}

\allowdisplaybreaks

\begin{document}

\title{
Heavy Majorana neutrino pair productions at the LHC \\ 
in minimal U(1) extended Standard Model
}
\author{Arindam Das}
\email{arindam@kias.re.kr}
\affiliation{
School of Physics, KIAS, Seoul 02455, Korea 
}
\author{Nobuchika Okada}
\email{okadan@ua.edu}
\author{Digesh Raut}
\email{draut@crimson.ua.edu}
\affiliation{
Department of Physics and Astronomy,
University of Alabama,
Tuscaloosa, AL 35487, USA
}

\vspace{2cm}

%\preprint{\today}
%=================================================================
\begin{abstract}
In our recent paper \cite{Das:2017flq} we explored a prospect of discovering the heavy Majorana right-handed neutrinos (RHNs) 
 at the future LHC in the context of the minimal non-exotic U(1) extended Standard Model (SM), where a pair of RHNs are created via decay of resonantly produced massive U(1) gauge boson ($Z^{\prime}$). 
We have pointed out that this model can yield a significant enhancement of the branching ratio of the $Z^\prime$ boson to a pair of RHNs, which is crucial for discovering the RHNs under the very severe LHC Run-2 constraint from the search for the $Z^\prime$ boson with dilepton final states.   
In this paper, we perform a general parameter scan to evaluate the maximum production rate of the same-sign dilepton final states 
 (smoking gun signature of Majorana RHNs production) at the LHC, while reproducing the neutrino oscillation data. 
We also consider the minimal non-exotic U(1) model with an alternative charge assignment. 
In this case, we find a further enhancement of the branching ratio of the $Z^\prime$ boson to a pair of RHNs compared to the conventional case, which opens up a possibility of discovering the RHNs even before the $Z^\prime$ boson at the future LHC experiment. 
\end{abstract}
%=================================================================

\maketitle
%\tableofcontents
%\clearpage
%=================================================================
%%%%%%%%%%%%%%%%%%%%%%%%%%%%%%%%%%%%%%%%%%%%%%%%
\section{Introduction}
\label{sec:1}
%%%%%%%%%%%%%%%%%%%%%%%%%%%%%%%%%%%%%%%%%%%%%%%%

The experimental evidence of the neutrino oscillation \cite{PDG} indicate that neutrinos have tiny but non-zero masses and flavor mixings. 
Since the neutrinos are massless in the Standard Model (SM), we need to extend the SM to incorporate the non-zero neutrino masses and flavor mixings.  
From the low energy effective theory viewpoint, we may introduce a dimension-5 operator \cite{Weinberg:1979sa} 
  involving the Higgs and lepton doublets, which violates the lepton number by two units. 
After the electroweak symmetry breaking, neutrinos acquire tiny Majorana masses 
  suppressed by the scale of the dimension-5 operator. 
For example, in a type-I seesaw  \cite{seesaw1,seesaw2,seesaw3,seesaw4,seesaw5}, 
heavy Majorana right-handed neutrinos (RHNs), which are singlet under the SM gauge group, are introduced, and the dimension-5 operator is generated by integrating them out.

If the RHNs have masses around 1 TeV or smaller, they can be produced at the  Large Hadron Collider (LHC). 
The same-sign dilepton in the final state, which indicates a violation of the lepton number, is a smoking-gun signature of RHN production. 
Since RHNs are singlet under the SM gauge group, they can be produced only through their mixings with the SM neutrinos. 
To reproduce the observed light neutrino mass scale, $m_\nu = \mathcal{O}(0.1)$ eV,  
  through the type-I seesaw mechanism with heavy neutrino masses at 1 TeV, a natural value 
  of the light-heavy neutrino mixing parameter is estimated to be $\mathcal{O} (10^{-6})$. 
With a general parametrization for the neutrino Dirac mass matrix \cite{Casas:2001sr}, this mixing parameter can be much larger. 
However, it turns out to be still small $\lesssim 0.01$ \cite{DO} in order to simultaneously satisfy a variety of experimental constraints, 
such as the neutrino oscillation data, the electroweak precision measurements and the lepton-flavor violating processes. 
Hence, the production rate of RHNs at the LHC is very suppressed.

In the simplest type-I seesaw scenario, the SM singlet RHNs are introduced only for the neutrino mass generation.  
The gauged $B-L$ extended SM \cite{mBL1, mBL2, mBL3, mBL4, mBL5, mBL6} may be a more compelling scenario, 
  which incorporates the type-I seesaw mechanism. 
In this model, the global U(1)$_{B-L}$ (baryon number minus lepton number) symmetry in the SM is gauged and the RHNs play the essential role to cancel the gauge and mixed-gravitational anomalies. 
After the spontaneous breaking of the $B-L$ symmetry, the RHNs acquire their Majorana masses, and the type-I seesaw mechanism is automatically implemented after the electroweak symmetry breaking. 
This model provides a new mechanism for the production of RHNs at the LHC. 
Since the $B-L$ gauge boson ($Z^\prime$) couples with both the SM fermions and the RHNs, once the $Z^\prime$ boson is resonantly produced at the LHC, its subsequent decay produces a pair of RHNs. 
Then, the RHNs decay into the SM particles through the light-heavy neutrino mixings: $N \to W^{\pm}  \ell^{\mp} $,  $Z \nu_{\ell}$, $Z {\overline \nu_\ell}$, $h \nu_{\ell} $, and  $h {\overline \nu_\ell} $.

Recently, in the context of the gauged $B-L$ models \cite{Kang:2015uoc, Yanagita:2017, Accomando:2017qcs},  
  the prospect of discovering the RHNs in the future LHC has been explored by simulation studies 
  of a resonant $Z^\prime$ boson production and its decay into a pair of RHNs. 
In Refs.~\cite{Kang:2015uoc, Accomando:2017qcs}, the authors have considered the trilepton final states, 
  $Z^\prime \to NN \to  \ell^{\pm} \ell^{\mp}\ell^{\mp}$ $\nu_{\ell}$ $ j j $.  
For example, in Ref.~\cite{Accomando:2017qcs} the signal-to-background ratio of $S/{\sqrt B} \simeq 10$ has been obtained 
   at the LHC with a 300 fb$^{-1}$ luminosity, 
   for the production cross section, $\sigma (pp \to Z^\prime \to NN \to  \ell^{\pm} \ell^{\mp}\ell^{\mp} \nu_{\ell} jj) = 0.37$ fb ($\ell = e$ or $\mu$),
   with the $Z^\prime$ and the RHN masses fixed as $m_{Z^\prime} = 4$ TeV and  $m_N = 400$ GeV, respectively. 
In Ref.~\cite{Yanagita:2017}, the authors have considered the final state with a same-sign dimuon and a boosted diboson, 
    $Z^\prime \to NN \to  \ell^{\pm} \ell^{\pm}$ $W^{\mp}W^{\mp}$ \footnote{For previous studies  of $Z^\prime \to NN \to  e^{\pm} \mu^{\mp}$ $W^{\pm}W^{\mp}$, see, for example, Ref.~\cite{Deppisch:2013cya}.}. 
For fixed masses, $m_{Z^\prime} = 3$ TeV and  $m_N = m_{Z^\prime}/4$, 
   they have obtained a cross section $\sigma (pp \to Z^\prime \to NN \to  \mu^{\pm} \mu^{\pm} W^{\mp}W^{\mp}) \simeq 0.1$ fb 
   for a 5$\sigma$ discovery at the LHC with a 300 fb$^{-1}$ luminosity.

Since the RHNs are produced from the $Z^\prime$ boson decay, 
  in exploring the future prospect of discovering the RHNs 
  we need to consider the current LHC bound on the $Z^\prime$ boson production, 
  which is already very severe.\footnote{
In Ref.~\cite{Yanagita:2017}, the authors have considered the $U(1)_{(B-L)_{3}}$ model \cite{Alonso:2017uky}, 
  in which only the third generation fermions couple to the $Z^\prime$ boson. 
  Hence, the current LHC bound on the $Z^\prime$ boson production is not applicable to the model, 
  although their simulation results, which we employ in this paper, are model-independent.
}
The primary mode for the $Z^\prime$ boson search at the LHC is via the dilepton final states, 
  $ pp \to Z^\prime \to \ell^+ \ell^-$ ($\ell=e$ or $\mu$).  
The current upper bound on the $Z^\prime$ boson production cross section times its branching ratio 
  into a lepton pair ($e^+e^-$ and $\mu^+ \mu^-$ combined) is given by $\sigma (pp \to Z^\prime \to \ell^+ \ell^-) \lesssim 0.2$ fb,  
  for $m_{Z^{\prime}}\gtrsim 3$ TeV at the LHC Run-2 with 36.1 fb$^{-1}$ luminosity \cite{ATLAS_Z_Search}. 
Since the number of SM background events is very small for such a high $Z^\prime$ boson mass region, 
   we naively scale the current bound (at $95 \%$ confidence level) to a future bound as 
\bea
\sigma (pp \to Z^\prime \to \ell^+ \ell^-)  \lesssim 0.2 \ {\rm fb}\times \frac{36.1}{\cal L}, 
\label{ZBound} 
\eea
where ${\cal L}$ (in units of fb$^{-1}$) is a luminosity at the future LHC. 
Here, we have assumed the worst case scenario, namely, 
  there is still no indication of the $Z^\prime$ boson production in the future LHC data. 
For example, at the High-Luminosity LHC with  ${\cal L} =300$ fb$^{-1}$, the bound becomes 
$\sigma (pp \to Z^\prime \to \ell^+ \ell^-) \lesssim 2.4\times 10^{-2}$ fb. 
Note that this value is much smaller than the RHN production cross section of ${\cal O}(0.1)$ fb obtained in the simulation studies. 
Taking into account the branching ratios $NN \to  \ell^{\pm} \ell^{\mp}\ell^{\mp} \nu_{\ell} jj$ 
  and $NN \to  \ell^{\pm} \ell^{\pm} W^{\mp}W^{\mp}$, 
  the original production cross section $\sigma (pp \to Z^\prime \to NN)$ must be rather large. 
Therefore, an enhancement of the branching ratio 
  ${\rm BR}(Z^\prime \to NN)$ over ${\rm BR}(Z^\prime \to \ell^+ \ell^-)$ 
  is crucial for the discovery of the RHNs in the future. 

In the worst case scenario with the $300$ fb$^{-1}$ luminosity, 
  we estimate an enhancement factor necessary to obtain 
   $\sigma (pp \to Z^\prime \to NN \to  \ell^{\pm} \ell^{\mp}\ell^{\mp} \nu_{\ell} jj)$ and
   $\sigma (pp \to Z^\prime \to NN \to  \mu^{\pm} \mu^{\pm}$ $W^{\mp}W^{\mp})= {\cal O}(0.1)$ fb,
   while $\sigma (pp \to Z^\prime \to \ell^+ \ell^-) \lesssim 2.4\times 10^{-2}$ fb. 
For $m_N \gg m_W = 80.4 $ GeV, $m_Z= 91.2$ GeV, and $m_h = 125.09$ GeV, we estimate the  branching ratios as 
  $\text{BR}(N\to W \ell) \simeq 0.5$ and $\text{BR}(N\to Z \nu) \simeq \text{BR}(N\to h \nu) \simeq 0.25$, 
   where we have considered one generation only. 
With $\text{BR}(W\to \ell \nu) \simeq 0.1$, $\text{BR}(W\to jj) \simeq 0.7$, $\text{BR}(Z \to \ell^+ \ell^-) \simeq 0.034$, $\text{BR}(Z \to \nu \nu) \simeq 0.2$, and $\text{BR}(Z \to jj) \simeq 0.7$, we estimate ${\rm BR}( N N \to \ell^+ \ell^-\ell^- \nu jj)={\rm BR}( N N \to \ell^- \ell^+\ell^+ \nu jj)\simeq 0.04$ and ${\rm BR}(N N \to \ell^\pm \ell^\pm W^\mp W^\mp) \simeq 0.125$. 
Hence, in order to obtain $\sigma (pp \to Z^\prime \to N N \to  \ell^{\pm} \ell^{\mp}\ell^{\mp}  \nu_{\ell} jj) \gtrsim 0.37$ fb \cite{Accomando:2017qcs} and $\sigma (pp \to Z^\prime \to NN \to  \ell^{\pm} \ell^{\pm}W^{\mp}W^{\mp} \gtrsim 0.1$ fb \cite{Yanagita:2017}, 
we find $\sigma (pp \to Z^\prime \to N N) \gtrsim 4.62$ fb and $0.8$ fb, respectively. 
Hence, the enhancement factors we need are 
\bea
 \frac{\text{BR}(Z^\prime \to N N)} {\text{BR}(Z^\prime \to \ell^+ \ell^-)} \gtrsim 192\;  \rm{and}\;  33.3, 
 \label{EFactor1}
\eea
respectively. 
Even for the same sign dilepton final states, we have found that a huge enhancement factor is required.  
Note that we only have $\frac{\text{BR}(Z^\prime \to NN)} {\text{BR}(Z^\prime \to \ell^+ \ell^-)} \simeq 0.5$ in the minimal $B-L$ model.

In this paper we consider a simple extension of the SM which can realize the branching ratio 
  $\text{BR}(Z^\prime \to NN) \gg \text{BR}(Z^\prime \to \ell^+ \ell^-)$.  
The model is based on the gauge group,  SU(3)$_c \times$SU(2)$_L \times$U(1)$_Y\times$U(1)$_X$, 
   where U(1)$_X$ is a generalization of U(1)$_{B-L}$ such that the U(1)$_{X}$ charges of particles are realized as a linear combination of the SM U(1)$_Y$ 
   and U(1)$_{B-L}$ charges (the so-called non-exotic U(1)$_X $ model \cite{Appelquist:2002mw}). 
Three generations of the RHNs are added to cancel the gauge and the gravitational anomalies. 
We consider two cases for the $B-L$ charge assignment for the RHNs: the conventional and the alternative cases. 
In the conventional case, a $B-L$ charge $-1$ is assigned to all three RHNs, while in the alternative case, 
  a $B-L$ charge $-4$ is assigned to two of the RHNs and $+5$ for the third one.

In our recent paper \cite{Das:2017flq}, we considered the minimal U(1)$_X$ model 
  with the conventional charge assignment and pointed out 
  that the model can yield a significant enhancement of the branching ratio of $Z^\prime$ boson to a pair of RHNs. 
We focused on the same-sign dimuon final state which is a smoking gun signature of Majorana RHNs production at the LHC. 
With such an enhancement and a realistic model-parameter choice to reproduce the neutrino oscillation data, 
  we concluded that the possibility of discovering the RHNs in the future implies 
  that the LHC experiments will discover the $Z^\prime$ boson well before the RHNs. 
In this paper, we extend the analysis in our previous paper and perform a general parameter scan 
  to evaluate the maximum production rate of the same-sign dilepton final state at the LHC, 
  while reproducing the neutrino oscillation data. 
We also consider the alternative charge assignment and find a huge enhancement of the branching ratio 
  of $Z^\prime$ boson to a pair of RHNs compared to the conventional case. 
Performing a general parameter scan for this case, we find a possibility of discovering the RHNs 
  even before the $Z^\prime$ boson at the future LHC experiments.

The paper is organized as follows. 
In the next section, we present the minimal U(1)$_X$ model with a conventional charge assignment. 
After considering the production of the RHNs, we discuss the prospect of discovering the RHNs 
  through their pair production from the decay of U(1)$_X$ gauge boson ($Z^\prime$) at the future LHC experiments. 
In Sec.~\ref{sec:3}, we present the minimal U(1)$_X$ model with an alternative charge assignment, 
  and discuss the prospect of discovering the RHNs in this case. 
In Sec.~\ref{sec:4}, we consider the RHN decay process in details and 
  employ a general parametrization for the neutrino Dirac mass matrix 
  to reproduce the neutrino oscillation data. 
Performing general parameter scans, we evaluate the maximum branching ratio
  into the signal process, $NN \to \ell^\pm  \ell^\pm W^\mp W^\mp$, 
  and discuss the prospect of discovering the RHN at the future LHC 
  in the minimal U(1)$_X$ model with both the conventional and the alternative charge assignments. 
Sec.~\ref{sec:5} is devoted to conclusions.

%%%%%%%%%%%%
\section{Minimal U(1)$_X$ Model}
\label{sec:2}
%%%%%%%%%%%%

%%%%%%%%%%%%%%%%%%%%%%%%%%%%%%%%%%%%%%%%%%%%%
\begin{table}[t]
\begin{center}
\begin{tabular}{c|ccc|c}
      &  SU(3)$_c$  & SU(2)$_L$ & U(1)$_Y$ & U(1)$_X$  \\ 
\hline
$q^{i}_{L}$ & {\bf 3 }    &  {\bf 2}         & $ 1/6$       & $(1/6) x_{H} + (1/3) x_{\Phi}$    \\
$u^{i}_{R}$ & {\bf 3 }    &  {\bf 1}         & $ 2/3$       & $(2/3) x_{H} + (1/3) x_{\Phi}$  \\
$d^{i}_{R}$ & {\bf 3 }    &  {\bf 1}         & $-1/3$       & $-(1/3) x_{H} + (1/3) x_{\Phi}$  \\
\hline
$\ell^{i}_{L}$ & {\bf 1 }    &  {\bf 2}         & $-1/2$       & $(-1/2) x_{H} - x_{\Phi}$  \\
$e^{i}_{R}$    & {\bf 1 }    &  {\bf 1}         & $-1$                   & $-x_{H} - x_{\Phi}$  \\
\hline
$H$            & {\bf 1 }    &  {\bf 2}         & $- 1/2$       & $(-1/2) x_{H}$  \\  
\hline
$N^{i}_{R}$    & {\bf 1 }    &  {\bf 1}         &$0$                    & $- x_{\Phi}$   \\

$\Phi$            & {\bf 1 }       &  {\bf 1}       &$ 0$                  & $ + 2x_{\Phi}$  \\ 
\hline
\end{tabular}
\end{center}
\caption{
Particle content of  the U(1)$_X$ model, where $i=1, 2, 3$ are generation indices.
Without loss of generality, we fix $x_{\Phi}=1$ throughout this paper. 
}
\label{tab1}
\end{table}  
%%%%%%%%%%%%%%%%%%%%%%%%%%%%%%%%%%%%%%%%%%%%%

We first consider the minimal U(1)$_X$ extension of SM.\footnote{
   In Refs.~\cite{Hashimoto:2014ela, Accomando:2016sge, Okada:2016tci,  Okada:2017cvy}, 
   a variety of phenomenology of the minimal U(1)$_X$ extended SM, 
   such as the electroweak vacuum stability, LHC physics, dark matter physics, 
   and cosmological inflation, have been extensively studied.} 
The model is based on the gauge group,  SU(3)$_c \times$SU(2)$_L\times$U(1)$_Y\times$U(1)$_X$, 
   where U(1)$_X$ is a generalization of U(1)$_{B-L}$ such that the U(1)$_X$ is a generalization of U(1)$_{B-L}$ such that the U(1)$_{X}$ charges of particles are realized as a linear combination of the SM U(1)$_Y$ 
   and U(1)$_{B-L}$ charges (the so-called non-exotic U(1)$_X$ model \cite{Appelquist:2002mw}). 
The structure of the theory is the same as the $B-L$ model except for a U(1)$_X$ charge assignment. 
The particle content is listed in Table~\ref{tab1}. 
In addition to the SM particle content, this model includes three generations of RHNs ($N^i_R$) 
  required for gauge anomaly cancellations, 
  a new Higgs field ($\Phi$) which breaks the U(1)$_X$ gague symmetry, and a  U(1)$_X$ gauge boson ($Z^\prime$). 
The U(1)$_X$ charges are defined in terms of two real parameters $x_{H}$ and $x_{\Phi}$, 
  which are the U(1)$_X$ charges associated with $H$ and $\Phi$, respectively. 
In this model, $x_{\Phi}$ always appears as a product with the U(1)$_{X}$ gauge coupling and is not an
   independent free parameter, which we fix to be $x_{\Phi}=1$ throughout this paper. 
Hence, U(1)$_X$ charges of the particles are defined by a single free parameter $x_H$. 
Note that this model is identical to the minimal $B-L$ model in the limit of $x_{H}=0$. 

In the minimal U(1)$_X$ model, the Yukawa sector of the SM is extended to include
\bea
\mathcal{L} _{Y}\supset -\sum_{i=1}^{3} \sum_{j=1}^{3} Y_{D}^{ij} \overline{\ell_{L}^{i}} H N_{R}^{j}-\frac{1}{2} \sum_{i=1}^{3} Y_{N}^{k} \Phi \overline{N_{R}^{k \ c}} N_{R}^{k}+ \rm{h. c.}, 
\label{U1XYukawa}
\eea 
where the first and second terms are the Dirac and Majorana Yukawa couplings. 
Here we use a diagonal basis for the Majorana Yukawa coupling without loss of generality. 
We assume a suitable Higgs potential for $\phi$ and $H$ to develop their vacuum expectation values, 
  ${v_\Phi}$ and $ {v_h} =246$ GeV, respectively. 
After the U(1)$_X$ and the electroweak symmetry breakings, U(1)$_X$ gauge boson mass, 
 the Majorana masses for the RHNs, and neutrino Dirac masses are generated:
\bea
  m_{Z^\prime} = g_X \sqrt{4 v_\Phi^2+  \frac{1}{4}x_H^2 v_h^2} \simeq 2 g_X v_\Phi , \; \;  m_{N^i}=\frac{Y_N^i}{\sqrt{2}} v_\Phi, \; \; m_{D}^{ij}=\frac{Y_{D}^{ij}}{\sqrt{2}} v_h,
  \eea   
where $g_X$ is the U(1)$_X$ gauge coupling, and 
  we have used the LEP constraint, ${v_\Phi}^2 \gg {v_h}^2$ \cite{Carena:2004xs, Heeck:2014zfa, DB}.

Let us now consider the RHN production via $Z^{\prime}$ boson decay. 
The $Z^{\prime}$ boson decay width into a pair of SM chiral fermions ($f_L$) is given by 
\bea
\Gamma({Z^{\prime}\to {\overline{f_L}} f_L})
 =\frac{N_c g_X^2 }{24 \pi} Q_{f_L}^2 m_{Z^{\prime}} \left(1-\frac{4m^2_f}{{m^2_{Z^\prime}}}\right)^{1/2}\left(1-2\frac{m^2_f}{{m^2_{Z^\prime}}}\right) \simeq \frac{N_c g_X^2 }{24 \pi} Q_{f_L}^2 m_{Z^{\prime}},
\label{Zwidths1}
\eea 
where $N_c = 1 (3)$ is the color factor for lepton (quark), $m_f$ ($ Q_{f_L}$) is the mass (charge) of the SM fermions, and we have used $m_{f_L}^2 \ll m^2_{Z^\prime}$ in the final expression. 
Similarly, the partial $Z^{\prime}$ boson decay width into a pair of single generation of Majorana RHNs is given by 
\bea
\Gamma(Z^{\prime}\to N N)
 = \frac{g_X^2 }{24\pi} Q^2_{N_R} m_{Z^{\prime}} \left(1-\frac{4 m^2_N}{{m^2_{Z^\prime}}}\right)^{3/2}, 
\label{Zwidths2}
\eea 
where, $m_N$ and $Q_{N_R}$ are the mass and the U(1)$_X$ charge of the RHN, respectively.

%%%%%%%%%%%%%%%%%%%%%%%%%%
%%%%%%%%%%%%%%%%%%%%%%%%%%%
\begin{figure}[t]
\begin{center}
\includegraphics[width=0.6\textwidth,angle=0,scale=0.80]{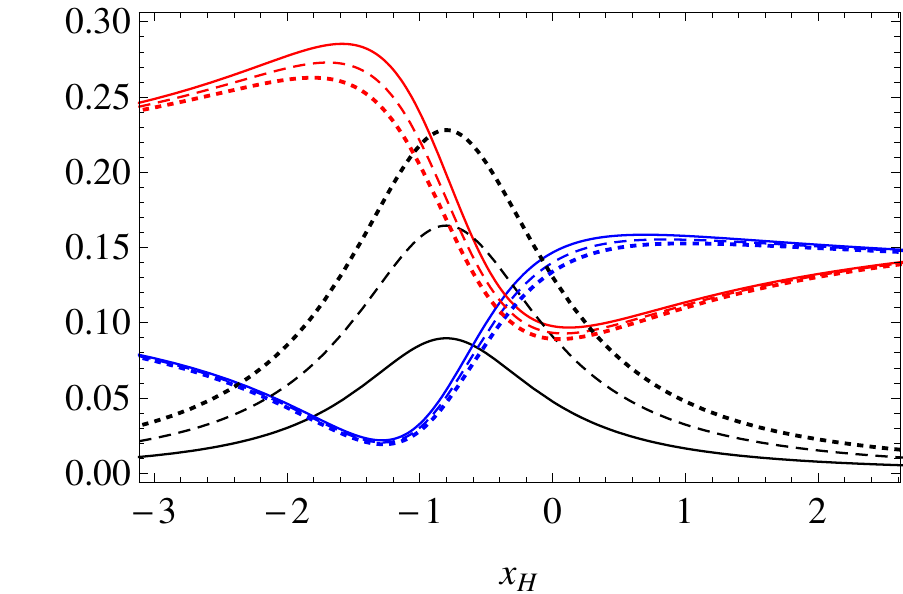} \; \;
\includegraphics[width=0.6\textwidth,angle=0,scale=0.80]{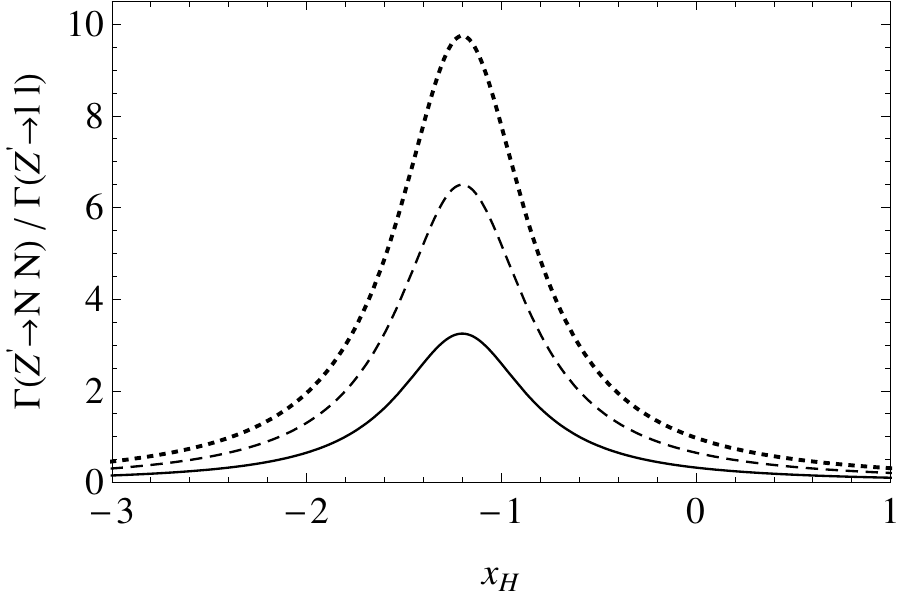}
\end{center}
\caption{
For the minimal U(1)$_X$ model, the left panel shows the branching ratios of $Z^{\prime}$ 
   as a function of $x_H$ with a fixed $m_{ Z^{\prime}} = 3$ TeV. 
The solid lines correspond to $m_{N^1} = m_{ Z^{\prime}}/4$ and $m_{N^{2,3}} > m_{ Z^{\prime}}/2$;
the dashed (dotted) lines correspond to $m_{N^{1,2}} = m_{ Z^{\prime}}/4$ and $m_{N^{3}} > m_{ Z^{\prime}}/2$ ($m_{N^{1,2,3}} = m_{ Z^{\prime}}/4$ ). 
From top to bottom, the solid (red, black and blue) lines at $x_H = -1$ are the branching ratios to the first generations of jets (up and down quarks), 
  RHNs, and charged leptons, respectively. 
The lines for the RHN final states correspond to the sum of the branching ratio to all possible RHNs. 
In the right panel, we show the ratio of the partial decay widths of $Z^\prime$ boson into RHNs and dilepton final states. 
The line codings are the same as in the left panel.
}
\label{Minimal}
\end{figure}
%%%%%%%%%%%%%%%%%%%%%%%%%%%
%%%%%%%%%%%%%%%%%%%%%%%%%%

In the left panel of Fig.~\ref{Minimal}, 
   we show the $Z^{\prime}$ branching ratios for the minimal U(1)$_X$ model with a fixed $m_{Z^\prime} = 3$ TeV. 
The solid lines correspond to $m_{N^1} =m_{ Z^{\prime}}/4 $ and $m_{N^{2,3}} > m_{ Z^{\prime}}/2$;
the dashed (dotted) line corresponds to $m_{N^{1,2}} = m_{ Z^{\prime}}/4$ and $m_{N^{3}} > m_{ Z^{\prime}}/2$ ($m_{N^{1,2,3}} = m_{ Z^{\prime}}/4$). 
For the SM final states, we show branching ratios to only the first generation dilepton and jets (sum of the jets from up and down quarks). 
The lines for the RHN final states correspond to the sum of the branching ratio to all possible RHNs. 
The plot shows the enhancement of the branching ratios into RHNs around $x_H = -0.8$, 
  with the maximum values of the branching ratios, $0.09$, $0.16$, and $0.23$, 
  for the cases with one, two, and three generations of RHNs, respectively. 
For the minimal $B-L$ model ($x_H = 0$), the branching ratios are only $0.05$, $0.09$, and $0.13$, respectively.

As discussed in Sec.~\ref{sec:1}, the discovery of RHNs at the collider via the $Z^\prime$ decay requires 
    some enhancement of the RHN production cross section, 
    because the LHC Run-2 results already set the very severe upper bound 
    on the $Z^\prime$ production cross section with the dilepton final states. 
To see how much enhancement can be achieved in the minimal U(1)$_X$ model, let us now consider 
   a ratio of the partial decay widths into a pair of $NN$ and dilepton final states, 
   which is nothing but the ratio of the $NN$ and dilepton production cross section.  
Using Eqs.~(\ref{Zwidths1}) and (\ref{Zwidths2}), this ratio is given by
\bea
\frac{\Gamma(Z^{\prime}\to N N)}{\Gamma({Z^{\prime}\to {\bar{\ell}} \ell})}
 = \frac{4 Q^2_{N_R}}{8 + 12 x_H + 5 x_H^2} \left(1-\frac{4 m^2_N}{{m^2_{Z^\prime}}}\right)^{3/2},  
\label{ZtoEE}
\eea 
for only one generation of RHNs and charged leptons in the final states.

In the right panel of Fig.~\ref{Minimal}, we show the ratio as a function of $x_H$. 
We find the peaks at $x_H= -1.2$ with the maximum values of $3.25$, $6.50$, and $9.75$, respectively.\footnote{
  In the left panel of Fig.~\ref{Minimal}, we can see that the branching ratio to the dijet final states 
  is also significantly enhanced. 
  As we have commented in Ref.~\cite{Das:2017flq}, 
  the LHC constraint on the $Z^\prime$ boson production cross section 
  with the dilepton final states is still stronger than 
  that with the dijet final states even with such an enhancement.  
} 
Although we have obtained remarkable enhancement factors, these are not large enough,  
   compared to the values required in the worst case scenario (see Eq.~(\ref{EFactor1})). 
Since the enhancement required for the trilepton final states is extremely large, 
   we focus on the same sign dilepton and diboson final states in the rest of this section. 

%%%%%%%%%%%%%%%%%%%%%%%%%
%%%%%%%%%%%%%%%%%%%%%%%%%
\begin{figure}[t]
\begin{center}
\includegraphics[width=0.8\textwidth,angle=0,scale=0.8]{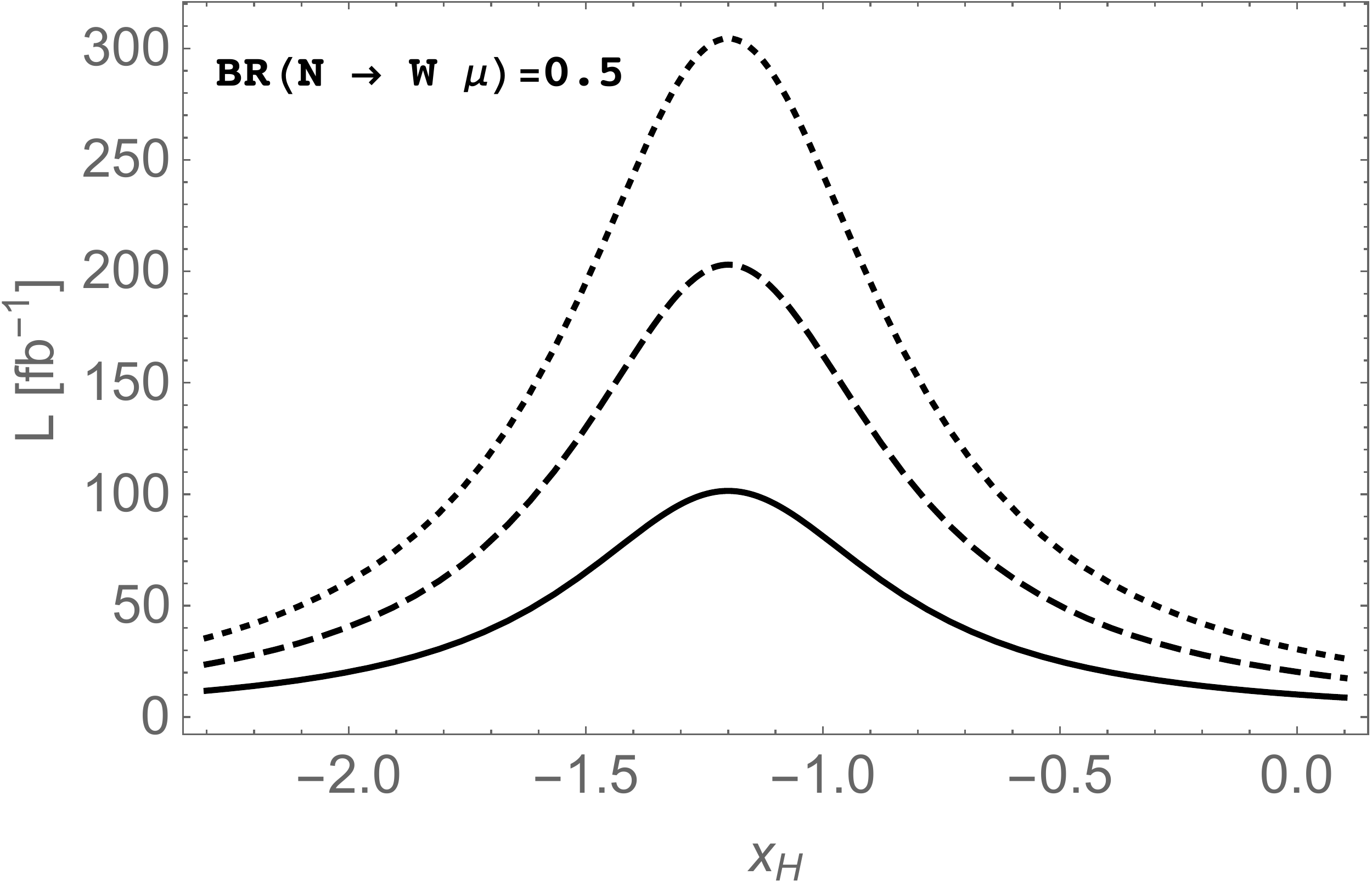}
\end{center}
\caption{
The plot shows the luminosity required to obtain 25 signal events of the $Z^\prime$ boson 
  as a function of $x_H$, for fixed values of $m_{ Z^{\prime}} = 3$ TeV and $\text{BR}(N\to W \mu) \simeq 0.5$. 
The solid lines correspond to $m_{N^1} = m_{ Z^{\prime}}/4$ and $m_{N^{2,3}} > m_{ Z^{\prime}}/2$;
the dashed (dotted) lines correspond to $m_{N^{1,2}} = m_{ Z^{\prime}}/4$ and $m_{N^{3}} > m_{ Z^{\prime}}/2$ ($m_{N^{1,2,3}} = m_{ Z^{\prime}}/4$ ).  
}
\label{DiscConv1}
\end{figure}
%%%%%%%%%%%%%%%
%%%%%%%%%%%%%%% 
Let us now consider an optimistic case and assume that the $Z^\prime$ boson has been discovered 
  at the LHC.
In this case, we remove the constraint $\sigma (pp \to Z^\prime \to \ell^+ \ell^-) \lesssim 2.4\times 10^{-2}$ fb. 
%Instead, we estimate the cross section $\sigma (pp \to Z^\prime \to \ell^+ \ell^-)$ in order to achieve the RHN production cross section of $\sigma (pp \to Z^\prime \to N N) \simeq 0.8$ fb  required for the $5\sigma$ discovery with the $300$ fb$^{-1}$ luminosity. 
According to \cite{Yanagita:2017}, the cross section required for the 5$\sigma$ discovery of the RHNs at the LHC with a 300 fb$^{-1}$ luminosity is $\sigma (pp \to Z^\prime \to NN \to  \mu^{\pm} \mu^{\pm} W^{\mp}W^{\mp}) \simeq 0.1$ fb. 
Although it is difficult for us to evaluate systematic errors, we here very naively 
require ad-hoc benchmark number of signal events to be 25 for the discovery of 
the $Z^\prime$ boson production, since the number of SM background events for a high $Z^\prime$ boson mass region ($m_{Z^\prime}\gtrsim 3$ TeV) is very small. 
Hence, we estimate the luminosity (${\mathcal L}$) for 25 signal events of the $Z^\prime$ boson production as follows: 
\bea
\frac{\sigma (pp \to Z^\prime \to NN \to   \mu^{\pm} \mu^{\pm}W^{\mp}W^{\mp})}{\sigma (pp \to Z^\prime \to \ell^+ \ell^-)}
\simeq  \frac{0.1}{\frac{25}{{\mathcal L}}}.   
\eea
For a degenerate mass spectrum for the RHNs, $\sigma (pp \to Z^\prime \to NN \to   \mu^{\pm} \mu^{\pm}W^{\mp}W^{\mp})  = \sigma (pp \to Z^\prime \to N_{m}^i N_m^i  ) \times \sum_{i}{\rm BR}(N_{m}^i N_m^i \rightarrow \mu^\pm \mu^\pm W^\mp W^\mp)$, and we obtain
\bea
\label{enhancement}
{\mathcal L} ({\rm fb}^{-1}) &\simeq& 250 \times \sum_{i=1}{\rm BR}(N_{m}^i N_m^i \rightarrow \mu^\pm \mu^\pm W^\mp W^\mp) \times \frac{\Gamma ( Z^\prime \to N_{m}^i N_m^i  )}{\Gamma(Z^\prime \to \ell^+ \ell^-)}, 
\eea 
where $\frac{\Gamma ( Z^\prime \to N_{m}^i N_m^i  )}{\Gamma(Z^\prime \to \ell^+ \ell^-)}$ is shown in the right panel of Fig.~\ref{Minimal}.

For the fixed values of $m_{ Z^{\prime}} = 3$ TeV and $\text{BR}(N\to W \mu) \simeq 0.5$, we employ  Eq.~(\ref{enhancement}) and show the luminosity (${\mathcal L}$) as a function of $x_H$ in Fig.~\ref{DiscConv1}. 
The solid lines correspond to $m_{N^1} = m_{ Z^{\prime}}/4$ and $m_{N^{2,3}} > m_{ Z^{\prime}}/2$, while 
the dashed (dotted) lines correspond to $m_{N^{1,2}} = m_{ Z^{\prime}}/4$ and $m_{N^{3}} > m_{ Z^{\prime}}/2$ ($m_{N^{1,2,3}} = m_{ Z^{\prime}}/4$ ). 
Hence, $x_H$ is constrained to be in the range of $-2\lesssim x_H \lesssim 0$. 
%Hence, the required luminosities will be reached at the future LHC for all the values of $x_H$. 
For example, let us consider the case of $x_H = -1.2$ for which the ratio  $\Gamma(Z^{\prime}\to N N)/\Gamma({Z^{\prime}\to {\bar{\ell}} \ell})$ reaches the maximum values of $3.25$, $6.50$, and $9.75$ for one, two, and three degenerate RHNs, respectively. 
Hence, we obtain the luminosities ${{\mathcal L}}({\rm fb}^{-1}) \simeq 102$, $203$ and $305$ for one, two and three generations of degenerate RHNs, respectively. 
These luminosities will be reached in the near future.

%%%%%%%%%%%%%%
\section{Alternative U(1)$_X$ model}
\label{sec:3}
%%%%%%%%%%%%%%

\begin{table}[t]
\begin{center}
\begin{tabular}{c|ccc|c}
      &  SU(3)$_c$  & SU(2)$_L$ & U(1)$_Y$ & U(1)$_{X}$  \\ 
\hline
$N^{1}_{R}$    & {\bf 1 }    &  {\bf 1}         &$0$                    & $- 4$   \\
$N^{2}_{R}$    & {\bf 1 }    &  {\bf 1}         &$0$                    & $- 4$   \\
$N^{3}_{R}$    & {\bf 1 }    &  {\bf 1}         &$0$                    & $5$   \\
\hline
$H_E$            & {\bf 1 }       &  {\bf 2}       &$ -\frac{1}{2}$                  & $ (-1/2) x_{H}+3$  \\ 
$\Phi_A$            & {\bf 1 }       &  {\bf 1}       &$ 0$                  & $ +8$  \\ 
$\Phi_B$            & {\bf 1 }       &  {\bf 1}       &$ 0$                  & $ -10$  \\ 
\end{tabular}
\end{center}
\caption{New particle content of the alternative U(1)$_X$ model.}
\label{tab2}
\end{table}

There is another way to assign the $B-L$ charges for the three RHNs to achieve gauge anomaly cancellations. 
The $B-L$ charge $-4$ is assigned to the first two generation of RHNs ($N^{1,2}$), while $-5$ for  $N^3$  \cite{Montero:2007cd}. 
In addition to the SM particle content, the new particle content of this ``alternative U(1)$_X$ model'' is listed in Table~\ref{tab2}. 
The U(1)$_X$ charge assignment for the SM particles is exactly the same as in the minimal U(1)$_X$ model.  
Here, we have introduced additional scalar fields, $H_E$ and $\Phi_{A,B}$.\footnote{
   One may consider an extended particle content (and some additional global symmetry) to forbid the seesaw mechanism 
   at the tree level and generate neutrino mass at the quantum levels \cite{Ma:2014qra}.
   } 
The new Higgs doublet $H_E$ generates the Dirac masses for the neutrinos, while the singlet scalars $\Phi_A$ and $\Phi_B$ generate Majorana masses for $N_R^{1,2}$ and $N^3_R$, respectively.

The Yukawa sector of the SM is extended to include 
\bea
\mathcal{L} _{Y}&\supset& -\sum_{i=1}^{3} \sum_{j=1}^{2} Y_{D}^{ij} \overline{\ell_{L}^{i}} H_E N_{R}^{j}-\frac{1}{2} \sum_{k=1}^{2} Y_{N}^{k} \Phi_A \overline{N_{R}^{k^{c}}} N_{R}^{k}+ \rm{h. c.}
\nonumber \\
& & -\frac{1}{2} Y_{N}^{3} \Phi_B \overline{N_{R}^{3^{c}}} N_{R}^{3}+ \rm{h. c.}. 
\label{ExoticYukawa}
\eea 
We assume a suitable scalar potential for $H$, $H_E$, $\Phi_{A}$, and $\Phi_B$, 
  in which these scalars develop their vacuum expectation values as follows:  
\bea
  \langle H \rangle =  \left(  \begin{array}{c}  
    \frac{1}{\sqrt{2}}v_h \\
    0 \end{array}
\right),   \;  \;  \; \; 
\langle H_E \rangle =  \left(  \begin{array}{c}  
    \frac{1}{\sqrt{2}} \tilde{v}_{h}\\
    0 \end{array}
\right),  \;  \;  \; \; 
\langle \Phi_A \rangle =  \frac{v_{A}}{\sqrt{2}},  \;  \;  \; \; 
\langle \Phi_B \rangle =  \frac{v_{B}}{\sqrt{2}}, 
\eea   
where we require that $v_h^2 + \tilde{v}^2_h = (246 \;  {\rm GeV})^2$. 
Associated with the U(1)$_X$ symmetry breaking, the RHNs and the U(1)$_X$ gauge boson ($Z^\prime$) 
  acquire their masses as 
\bea
 m_N^{1,2}&=&\frac{Y_N^{1,2}}{\sqrt{2}} v_A,  \; \;  \; \;  \; \;
 m_N^{3}=\frac{Y_N^{3}}{\sqrt{2}} v_B,  \nonumber \\ 
 m_{Z^\prime} &=& g_X \sqrt{64 v_{A}^2+ 100 v_{B}^2+  \frac{1}{4} x_H^2 v_h^2 + \left(-\frac{1}{2} x_H +3\right)^2 {\tilde v}_h^2} \nonumber \\
 &\simeq& g_X \sqrt{64 v_{A}^2+ 100 v_{B}^2} . 
\eea 
After the electroweak symmetry breaking, the neutrino Dirac masses, 
\bea
m_{D}^{ij}=\frac{Y_{D}^{ij}}{\sqrt{2}} \tilde{v}_h, 
\eea
are generated, and hence the seesaw mechanism is automatically implemented.

%%%%%%%%%%%%%%%%%%%%%%%%%%
%%%%%%%%%%%%%%%%%%%%%%%%%%%
\begin{figure}[t]
\begin{center}
\includegraphics[width=0.6\textwidth,angle=0,scale=0.8]{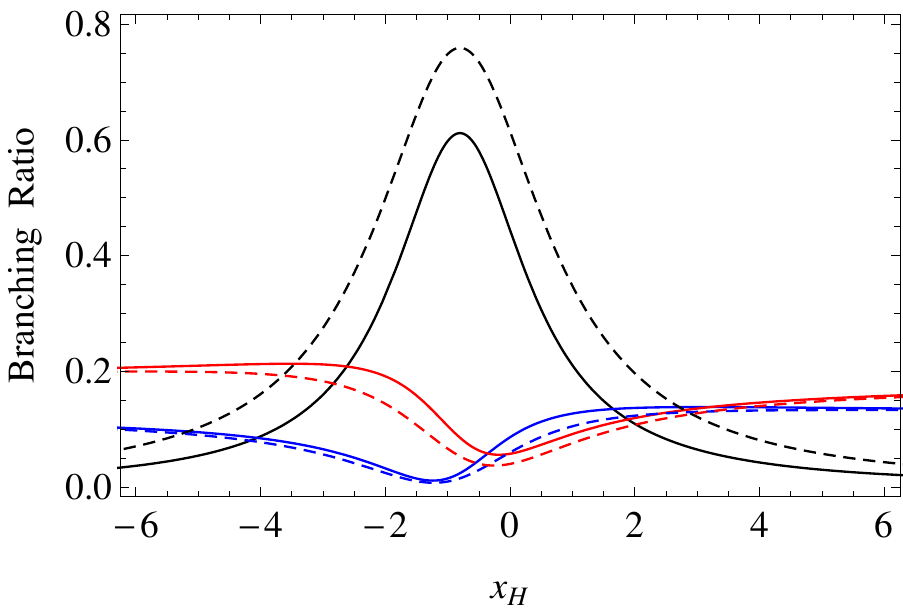}\; \;
\includegraphics[width=0.6\textwidth,angle=0,scale=0.8]{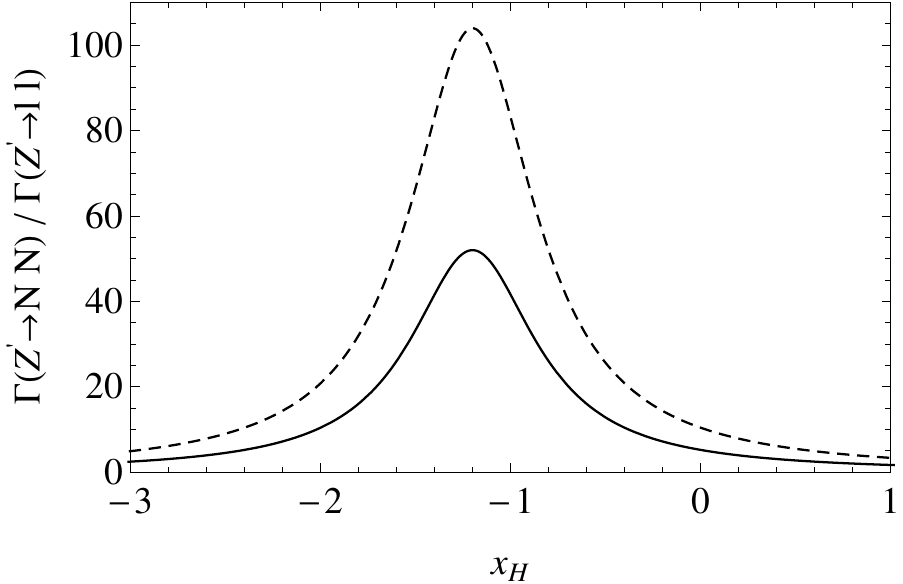}
\end{center}
\caption{For the alternative U(1)$_X$ model, the left panel shows the  branching ratios of $Z^{\prime}$ 
  as a function of $x_H$ with a fixed $m_{ Z^{\prime}} = 3$ TeV. 
The solid lines correspond to $m_{N^1} = m_{ Z^{\prime}}/4$ and $m_{N^{2}} > m_{ Z^{\prime}}/2$, 
   and the dashed lines correspond to $m_{N^{1,2}} = m_{ Z^{\prime}}/4$. 
From top to bottom, the solid (red, black and blue) lines at $x_H = -1$ are the branching ratios to the first generations of jets (up and down quarks), 
  RHNs, and charged leptons, respectively. 
The lines for the RHN final states correspond to the sum of the branching ratio to all possible RHNs.
In the right panel, we show the ratio of the partial decay widths of $Z^\prime$ boson into RHNs and dilepton final states. 
The line codings are the same as in the left panel.}
\label{Alternative}
\end{figure}
%%%%%%%%%%%%%%%
%%%%%%%%%%%%%%%

Let us now consider the branching ratios for  $Z^\prime$ decay. Note that in the alternative U(1)$_X$ model, 
     the charge assignment ensures the stability of $N_R^3$ and it is naturally a dark matter (DM) candidate \cite{OOR}. 
We may consider the scenario where the DM particle $N^3$ mainly communicates with the SM sector 
  via $Z^\prime$ boson exchange ($Z^\prime$ portal DM). 
In this case, we expect that the relic abundance constraint leads to $m_N^3 \simeq m_{Z^\prime}/2$. 
In the following, we consider this case and the partial decay width of the $Z^\prime$ into $N^3$ is neglected. 
The $Z^{\prime}$ boson decay width formulas are given by Eqs.~(\ref{Zwidths1}) and (\ref{Zwidths2}). 
In the alternative U(1)$_X$ model, $Q_{N_R} = - 4$  for $N_R^{1,2}$ in Eq.~(\ref{Zwidths2}).

For the alternative U(1)$_X$ model with a fixed $m_{ Z^{\prime}} = 3$ TeV, 
  we show the $Z^{\prime}$ branching ratios In the left panel of Fig.~\ref{Alternative}.   
The solid lines correspond to $m_{N^1} = m_{ Z^{\prime}}/4$ and $m_{N^{2,}} > m_{ Z^{\prime}}/2$. 
The dashed lines correspond to $m_{N^{1,2}} = m_{ Z^{\prime}}/4$. 
For the SM final states, we show branching ratios to only the first generation dilepton and jets (sum of the jets from up and down quarks). 
The lines for the RHN final states correspond to the sum of the branching ratio to all possible RHNs. 
The plot shows the enhancement of RHNs branching ratios around $x_H = -0.8$, 
  with the maximum values of the branching ratios, $0.612$ and $0.760$, 
  for the cases with one and two generations of RHNs, respectively. 
Note that even for the $B-L$ limit ($x_H = 0$),  the branching ratios are remarkably enhanced, $0.444$ and $0.615$, 
   compared to those obtained for the conventional charge assignment, $0.05$ and $0.09$, respectively.

In the right panel, we show the ratio of the partial decay widths into a pair of $NN$ and dilepton final states (see Eq.~(\ref{ZtoEE})). 
For U(1)$_X$ model with alternative charge assignment, we find the peaks in the ratio at $x_H= -1.2$, with the maximum values of $52.0$ and $104$, respectively. 
Note that even for the $B-L$ limit ($x_H = 0$), we have significant enhancements for the ratios of the partial decay widths 
   with the maximum values of $5.20$ and $10.4$, respectively, compared to $0.5$ for the conventional charge assignment. 
The maximum values of the enhancement factor for $x_H=-1.2$ are sufficiently large  
   for the RHN discovery with a same-sign dimuon and a boosted diboson final state (see Eq.~(\ref{EFactor1})).

%%%%%%%%%%%%%%%%%%%%%%%%%
%%%%%%%%%%%%%%%%%%%%%%%%%
\begin{figure}[t]
\begin{center}
\includegraphics[width=0.8\textwidth,angle=0,scale=0.8]{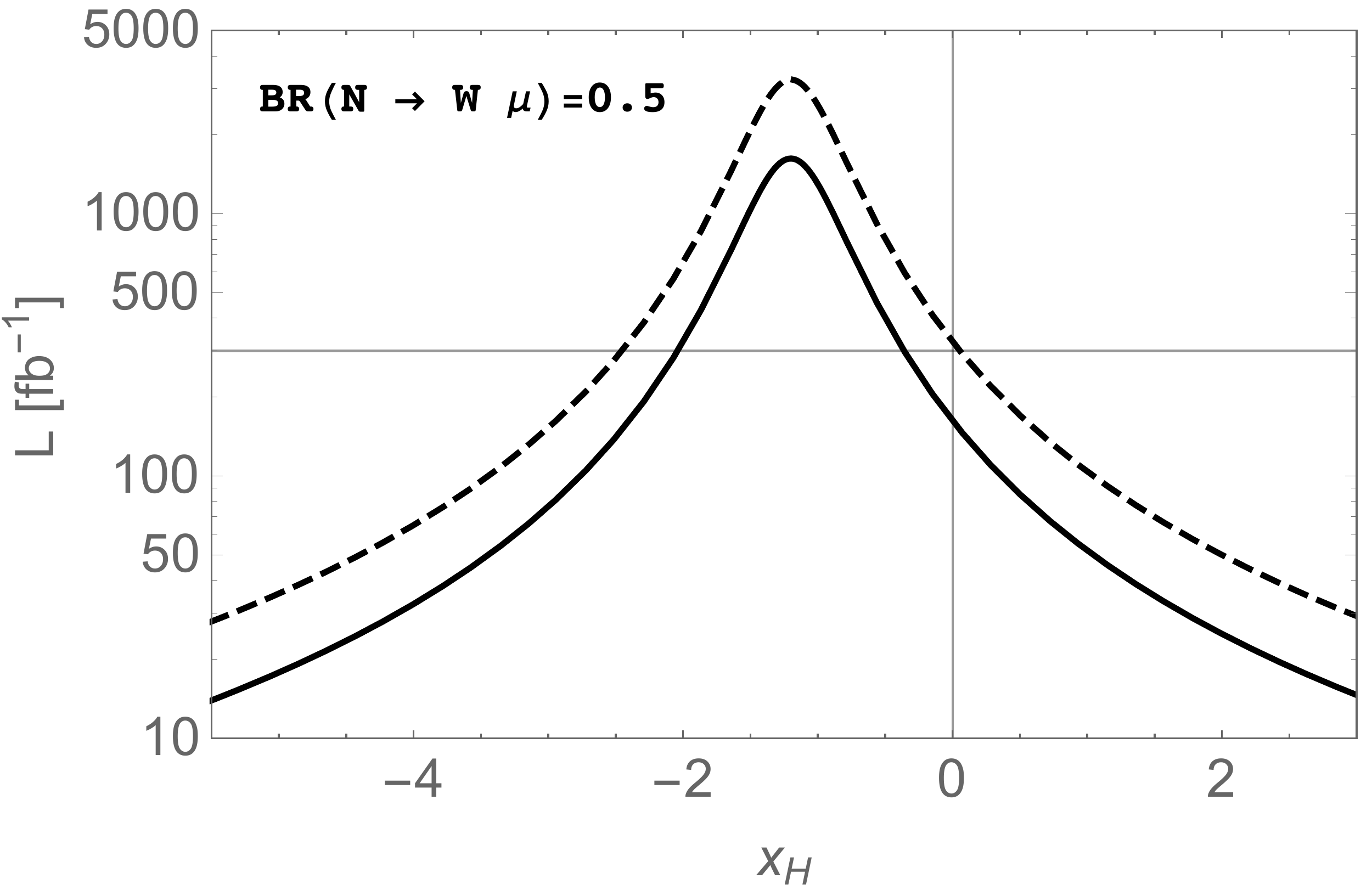}
\end{center}
\caption{The plot shows the luminosity required to obtain 25 signal events of  the $Z^\prime$ boson as a function of $x_H$, for fixed values of the $m_{ Z^{\prime}} = 3$ TeV and $\text{BR}(N\to W \mu) \simeq 0.5$. 
The solid line corresponds to $m_{N^1} = m_{ Z^{\prime}}/4$ and $m_{N^{2}} > m_{ Z^{\prime}}/2$, while the dashed line corresponds to $m_{N^{1,2}} = m_{ Z^{\prime}}/4$. 
The vertical solid line marks the $B-L$ limit ($x_H = 0$). The solid horizontal line corresponds to a luminosity value of 300 ${\rm fb}^{-1}$ required for the discovery of RHNs at the future LHC with a dimuon and a diboson final states.
}
\label{DiscAlt1}
\end{figure}
%%%%%%%%%%%%%%%
%%%%%%%%%%%%%%% 
Let us now consider the luminosity required for 25 signal events of the $Z^\prime$ boson production.  
For fixed values of the $m_{ Z^{\prime}} = 3$ TeV and $\text{BR}(N\to W \mu) \simeq 0.5$, we employ Eq.~(\ref{enhancement}) and show the luminosity (${\mathcal L}$) as a function of $x_H$ in Fig.~\ref{DiscAlt1}. 
The solid line corresponds to $m_{N^1} = m_{ Z^{\prime}}/4$ and $m_{N^{2}} > m_{ Z^{\prime}}/2$, while the dashed line corresponds to $m_{N^{1,2}} = m_{ Z^{\prime}}/4$. 
The vertical solid line marks the $B-L$ limit ($x_H = 0$).
Hence, $x_H$ is constrained to be in the range of $-4\lesssim x_H \lesssim2$. 
For example, for $x_H = -1.2$, the  luminosities for 25 signal events of the $Z^\prime$ boson production are found to be ${{\mathcal L}}({\rm fb}^{-1}) \simeq 1624$ and $3248$, for the case with one and two generation of degenerate RHNs, respectively. 
For the $B-L$ limit ($x_H = 0$) case,  the corresponding luminosities are ${{\mathcal L}}({\rm fb}^{-1}) \simeq 162$ and $325$,  for the case with one and two generation of degenerate RHNs, respectively. 
Interestingly, these values are comparable to the luminosities in the conventional case with the maximal enhancement ($x_H = -1.2$) for two and three generations of degenerate RHNs, respectively. 
The solid horizontal line denotes luminosity value of 300 ${\rm fb}^{-1}$ required for the discovery of RHNs at the future LHC with a dimuon and a diboson final states. 
Hence for example, for the case with two degenerate RHNs, Fig.~\ref{DiscAlt1} indicates that the RHNs will be discovered before the $Z^\prime$ boson for $-2.4 \lesssim x_H\lesssim 0$.

%%%%%%%%%%%%%%%%%%%%
\section{Realistic heavy neutrino branching ratios}
\label{sec:4}
%%%%%%%%%%%%%%%%%%%%

In the above analysis and the simulation studies, $\text{BR}(N\to W \mu) \simeq 0.5$ is assumed. 
However, note that in a realistic scenario to reproduce the neutrino oscillation data, 
  this branching ratio is smaller, which implies that more enhancement is required 
  to obtain a sufficient number of signal events. 
In this section we consider the RHN decay processes in more details. 

In the following analysis, we consider the case with degenerate RHNs, for simplicity. 
Using the Dirac and Majorana mass terms in Eqs.~(\ref{U1XYukawa}) or (\ref{ExoticYukawa}), 
   the neutrino mass matrix is expressed as  
\bea
{\cal M}_{\nu}=\begin{pmatrix}
0&&m_{D}\\
(m_{D})^{T}&&M_{N}
\end{pmatrix},
\label{typeInu}
\eea
where $m_D$ and $M_N$ are the Dirac and the Majorana mass matrices. 
Assuming the hierarchy of $|m_D^{ij} / m_N^j| \ll 1$, we have the seesaw formula for the light Majorana neutrinos as 
\bea
m_{\nu} \simeq - m_{D}(M_{N})^{-1}m_{D}^{T}. 
\label{seesawI}
\eea 
We express the light neutrino flavor eigenstate $(\nu)$ in terms of the mass eigenstates 
  of the light $(\nu_m)$ and heavy $(N_m)$ Majorana neutrinos such as 
$\nu \simeq \mathcal{N} \nu_m+\mathcal{R} N_m$, 
  where $\mathcal{R} =m_D (M_N)^{-1}$, $\mathcal{N}=\Big(1-\frac{1}{2}\mathcal{R}^\ast\mathcal{R}^T\Big)U_{\rm{MNS}}\simeq U_{\rm{MNS}}$, and $U_{\rm{MNS}}$ is the neutrino mixing matrix by which $m_\nu$ is diagonalized as 
\bea
  U_{\rm MNS}^T m_\nu U_{\rm MNS}  = D_\nu = {\rm diag}(m_1, m_2, m_3). 
\label{seesawII}
\eea 
In terms of the neutrino mass eigenstates, the charged current interaction is given by 
\bea 
\mathcal{L}_{CC}= 
 -\frac{g}{\sqrt{2}} W_{\mu}
  \overline{\ell_\alpha} \gamma^{\mu} P_L 
   \left( {\cal N}_{\alpha i} \nu_{m}^i+ {\cal R}_{\alpha i} N_{m}^i \right) + \rm{h.c.}, 
\label{CC}
\eea
where $\ell_\alpha$ are the three generations of the charged SM leptons, and $P_L =  (1- \gamma_5)/2$. 
Similarly, the neutral current interaction is given by 
\bea 
\mathcal{L}_{NC}&=& 
 -\frac{g}{2 \cos \theta_{\rm W}}  Z_{\mu} 
\Big[ 
  \overline{\nu_{m}^i} \gamma^{\mu} P_L ({\cal N}^\dagger {\cal N})_{ij} \nu_{m}^j
 +  \overline{N_{m}^i} \gamma^{\mu} P_L ({\cal R}^\dagger {\cal R})_{ij} N_{m}^j \nonumber \\
&+& \Big\{ 
  \overline{\nu_{m}^i} \gamma^{\mu} P_L ({\cal N}^\dagger  {\cal R})_{ij} N_{m}^j 
  + \rm{H.c.} \Big\} 
\Big] , 
\label{NC}
\eea
where $\theta_{\rm W}$ is the weak mixing angle.

The elements of the matrix ${\cal R}$ are arranged to reproduce the neutrino oscillation data, for which we adopt  
  $\sin^{2}2{\theta_{13}}=0.092$ \cite{Neut4} 
  along with
 $\sin^2 2\theta_{12}=0.87$, $\sin^2 2\theta_{23}=1.0$, 
 $\Delta m_{12}^2 = m_2^2-m_1^2 = 7.6 \times 10^{-5}$ eV$^2$, 
 and $\Delta m_{23}^2= |m_3^2-m_2^2|=2.4 \times 10^{-3}$ eV$^2$ \cite{PDG}. 
The neutrino mixing matrix is given by 
\bea
U_{\rm{MNS}} = \begin{pmatrix} c_{12} c_{13}&c_{12}c_{13}&s_{13}e^{-i\delta}\\-s_{12}c_{23}-c_{12}s_{23}s_{13}e^{i\delta}&c_{12}c_{23}-s_{12}s_{23}s_{13}e^{i\delta}&s_{23} c_{13}\\ s_{12}c_{23}-c_{12}c_{23}s_{13}e^{i\delta}&-c_{12}s_{23}-s_{12}c_{23}s_{13}e^{i\delta}&c_{23}c_{13}
\end{pmatrix} 
\begin{pmatrix}
1&0&0\\
0&e^{-i \rho_1}&0\\
0&0&e^{-i \rho_2}
 \end{pmatrix},
\eea
where $c_{ij}=\cos\theta_{ij}$,  $s_{ij}=\sin\theta_{ij}$, and $\rho_1$ and $\rho_2$ are the Majorana phases,\footnote{
 In the case with only two generations of RHNs, $\rho_2 = 0$. 
 } which are taken to be free parameters.   
Motivated by the recent measurement of the Dirac $CP$-phase, we set $\delta=\frac{3\pi}{2}$ \cite{CP-phase}.

From Eqs.~(\ref{seesawI}) and (\ref{seesawII}), we parameterize the Dirac mass matrix as \cite{Casas:2001sr}
\bea 
  m_D =  U_{\rm{MNS}}^* \sqrt{D_{\nu}} \; O \sqrt{M_N} ,  
\label{mD}
\eea
where $M_N$ is a diagonal matrix for the mass eigenvalues of the RHNs and $\sqrt{M_N}$ is defined 
   as a matrix with each element of $M_N$ square rooted, $O$ is a general orthogonal matrix, 
   and the matrix $\sqrt{D_\nu}$ will be defined later. 
For the light neutrino mass spectrum, we consider both the normal hierarchy (NH), $m_1< m_2< m_3$, 
   and the inverted hierarchy (IH), $m_3< m_1< m_2$.

Through its mixing with the SM leptons, a heavy neutrino mass eigenstate $N_m^i$ ($i=1,2,3$) decays 
   into $\ell W$, $\nu_{\ell} Z$, and $\nu_{\ell} h$ with the corresponding partial decay widths: 
\bea
\Gamma(N_m^i \rightarrow \ell_{\alpha} W)
 &=& \frac{1}{16 \pi} 
 \frac{ (M_{N}^2 - m_W^2)^2 (M_{N}^2+2 m_W^2)}{M_{N}^3 v_h^2} \times |R_{\alpha i}|^{2},
\nonumber \\
\Gamma(N_m^i \rightarrow \nu_{\ell_{\alpha}} Z)
 &=& \frac{1}{32 \pi} 
 \frac{ (M_{N}^2 - m_Z^2)^2 (M_{N}^2+2 m_Z^2)}{M_{N}^3 v_h^2} \times |R_{\alpha i}|^{2},
\nonumber \\
\Gamma(N_m^i \rightarrow \nu_{\ell_{\alpha}} h)
 &=& \frac{1}{32 \pi}\frac{(M_{N}^2-m_h^2)^2}{M_{N} v_h^2}\times |R_{\alpha i}|^{2},
\label{widths}
\eea 
where 
\bea
R_{\alpha i} =  (m_D)_{\alpha i} (M_N)^{-1}= U_{\rm{MNS}}^* \sqrt{D_{\nu}} \; O \sqrt{M_N} (M_N)^{-1} .  
\label{RMatrix}
\eea

%%%%%%%%%%%%%%%%%
\subsection{Minimal U(1)$_X$ model}
%%%%%%%%%%%%%%%%%
%%%%%%%%%%%%%%%%%%%%%%%%%
%%%%%%%%%%%%%%%%%%%%%%%%%
\begin{figure}[h]
\begin{center}
\includegraphics[width=0.6\textwidth,angle=0,scale=0.8]{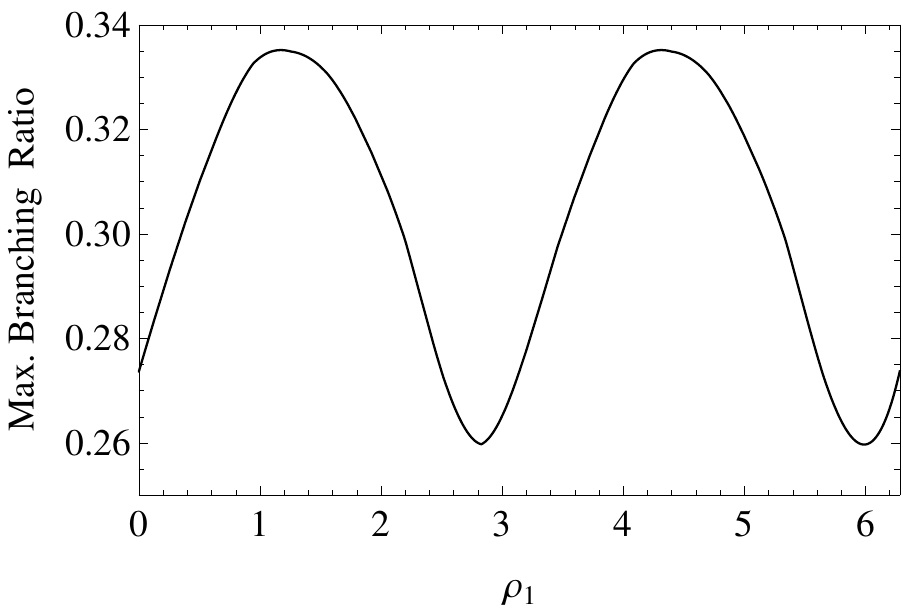}\; \;
\includegraphics[width=0.6\textwidth,angle=0,scale=0.8]{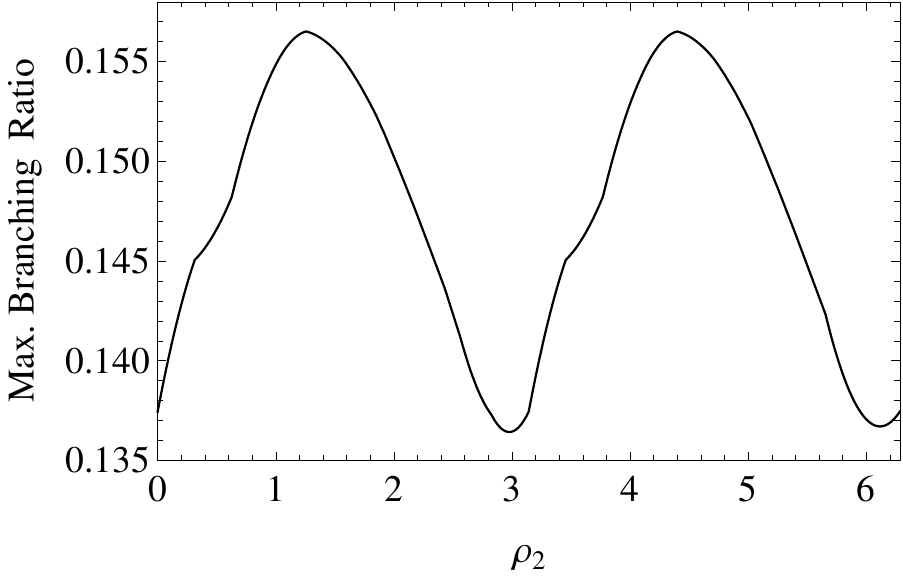}\; \; 
\end{center}
\caption{In the left (right) panel,  we show the prameter scan results for the maximum allowed branching ratios, $\sum_{i=1}^3{\rm BR}(N_{m}^i N_m^i \rightarrow \mu^\pm \mu^\pm W^\mp W^\mp)$, as a function of a Majorana phase $\rho_1$ ($\rho_2$) for the NH (IH) case. The solid curve denotes the maximum value of the branching ratio obtained after  performing a parameter scan for rest of the free parameters, $\theta_{1,2,3}, Y$, and $\rho_2 (\rho_1)$. From the figure we read the maximum value to be $0.337$ ($0.157$) for the NH (IH) case. 
}
\label{scan1}
\end{figure}
%%%%%%%%%%%%%%%
%%%%%%%%%%%%%%%

We first consider the minimal U(1)$_X$ model with three RHNs. 
In order to make our discussion simple, we assume the degeneracy of the heavy neutrinos 
   in mass such as $m_N= m_N^1=m_N^2 = m_N^3$. 
Here, for simplicity, we fix the lightest neutrino mass eigenvalue as $m_{\rm lightest} = 0.1 \times \sqrt{\Delta m_{12}^2}$, 
   by which the elements of the matrix $\sqrt{D_\nu} \equiv {\rm diag}(\sqrt{m_1},\sqrt{m_2},\sqrt{m_3})$ 
   are uniquely fixed for both NH and IH cases.    
We parameterize the general orthogonal matrix $O$ as 
\bea
O =
\begin{pmatrix}
1&0&0\\
0&\cos \theta_1& \sin \theta_1 \\
0&-\sin \theta_1& \cos \theta_1 \\
\end{pmatrix}
\begin{pmatrix}
\cos \theta_2& 0 &\sin \theta_2 \\
0&1&0 \\
-\sin \theta_2&0&\cos \theta_2 \\
\end{pmatrix} 
\begin{pmatrix}
\cos \theta_3& \sin \theta_3 &0 \\
-\sin \theta_3& \cos \theta_3&0 \\
0&0&1
\end{pmatrix}, 
\eea 
where $\theta_1, \theta_2$, and $\theta_3$ are complex numbers. 
With the inputs of the neutrino oscillation data and $M_N=m_{Z^\prime}/4$ with $m_{Z^\prime}=3$ TeV, 
  we have performed a scan for the free parameters ($\theta_1, \theta_2, \theta_3, \rho_1$, and $\rho_2$), 
  and found the maximum values of the branching ratio as 
$\sum_{i=1}^3{\rm BR}(N_{m}^i N_m^i \rightarrow \mu^\pm \mu^\pm W^\mp W^\mp) \simeq  0.337$ and $0.157$, 
  for the NH and IH cases, respectively (see Fig.~\ref{scan1}). 
In the analysis of Ref.~\cite{Das:2017flq}, the orthogonal matrix in Eq.~(\ref{RMatrix}) is taken to be a unit matrix, and the branching ratios have been found to be $\sum_{i=1}^3{\rm BR}(N_{m}^i N_m^i \rightarrow \mu^\pm \mu^\pm W^\mp W^\mp) \simeq  0.210$ and $0.154$, for the NH and IH cases, respectively. 
Thus, a general parameter scan yields a larger branching ratios. 
The branching ratio for the NH case is almost twice as large, while the IH case is almost the same as before. 

%%%%%%%%%%%%%%%%%%%%%%%%%
%%%%%%%%%%%%%%%%%%%%%%%%%
\begin{figure}[h]
\begin{center}
\includegraphics[width=0.8\textwidth,angle=0,scale=0.8]{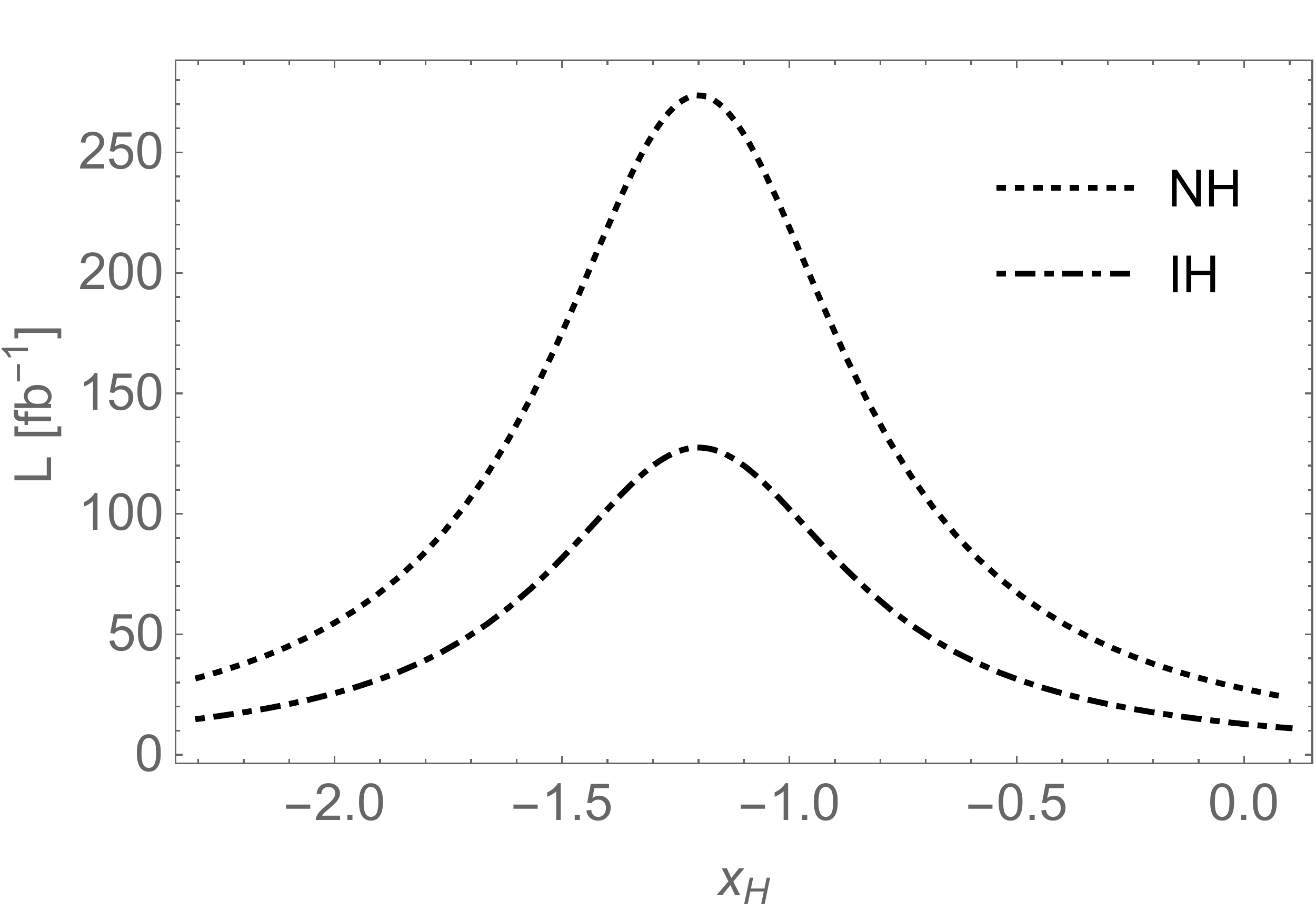}
\end{center}
\caption{The luminosity required to  obtain 25 signal events of  the $Z^\prime$ boson 
as a function of $x_H$ for fixed values of $m_{ Z^{\prime}} = 3$ TeV and  $m_{N^{1,2,3}} = m_{ Z^{\prime}}/4$. 
The dotted (dot-dashed) lines correspond to the NH (IH) case with $\sum_{i=1}^3{\rm BR}(N_{m}^i N_m^i \rightarrow \mu^\pm \mu^\pm W^\mp W^\mp) =0.337$ ($0.157$). 
}
\label{DiscConv2}
\end{figure}
%%%%%%%%%%%%%%%
%%%%%%%%%%%%%%%

Using these realistic values for branching ratios to reproduce the neutrino oscillation data, we now re-evaluate the luminosity required for 25 signal events of the $Z^\prime$ boson production. 
For fixed values of $m_{ Z^{\prime}} = 3$ TeV and $m_{N^{1,2,3}} = m_{ Z^{\prime}}/4$, we show the required luminosity as a function of $x_H$ in Fig.~\ref{DiscConv2}. The dotted (dot-dashed) lines correspond to the NH (IH) case. 
For three degenerate RHNs and for fixed values of $x_H = -1.2$ and $\text{BR}(N\to W \mu) \simeq 0.5$, we previously obtained the required luminosity to be $ {{\mathcal L}}({\rm fb}^{-1})  \simeq  305$. 
Using the realistic branching ratios for the RHNs, $\sum_{i=1}^3{\rm BR}(N_{m}^i N_m^i \rightarrow \mu^\pm \mu^\pm W^\mp W^\mp) =0.337$ and  $0.157$, the required luminosities are corrected to be ${{\mathcal L}}({\rm fb}^{-1})  \simeq 274$ and $128$ for the NH and the IH cases, respectively. 
Hence, for the realistic case, the required luminosity are reduced compared to the case of $\text{BR}(N\to W \mu) \simeq 0.5$.  
Accordingly, the allowed range of $x_H$ values for the NH (IH) case is reduced to be  $-2.3 \leq x_H \leq -0.16$ ($-1.9 \leq x_H \leq -0.54$).

If there is no indication of the $Z^\prime$ boson production at the future LHC with a dilepton final state, we obtain an upper bound on the $U(1)_X$ gauge coupling for a fixed $x_H$ value and the $Z^\prime$ boson mass. Using a narrow decay width approximation, the total production cross section of the $Z^\prime$ boson is proportional to  $\alpha_X = g_X^2/(4\pi)$. 
We refer the results in Refs.~\cite{Okada:2017dqs, Klasen:2016qux} for the upper bound $\alpha_X \lesssim 0.01$ \footnote{ When the $Z^\prime$ boson can decay into a pair of RHNs, the current LHC bound becomes slightly weaker \cite{Dev:2016xcp}.} for $x_H=-1.2$ and $m_{Z^\prime}= 3$ TeV from the ATLAS results with ${\cal L} = 36.1$ fb$^{-1}$. 
The upper bound on $\alpha_X$ scales as  
\bea
\alpha_X \lesssim 0.01\times \frac{36.1}{\cal L},
\eea
where ${\cal L}$ in units of fb$^{-1}$ is a luminosity at the future LHC.

%%%%%%%%%%%%%%%%%%%%%%
\subsection{Alternative U(1)$_X$ model}
%%%%%%%%%%%%%%%%%%%%%
%%%%%%%%%%%%%%%%
%%%%%%%%%%%%%%%%
\begin{figure}[h]
\begin{center}
\includegraphics[width=0.6\textwidth,angle=0,scale=0.8]{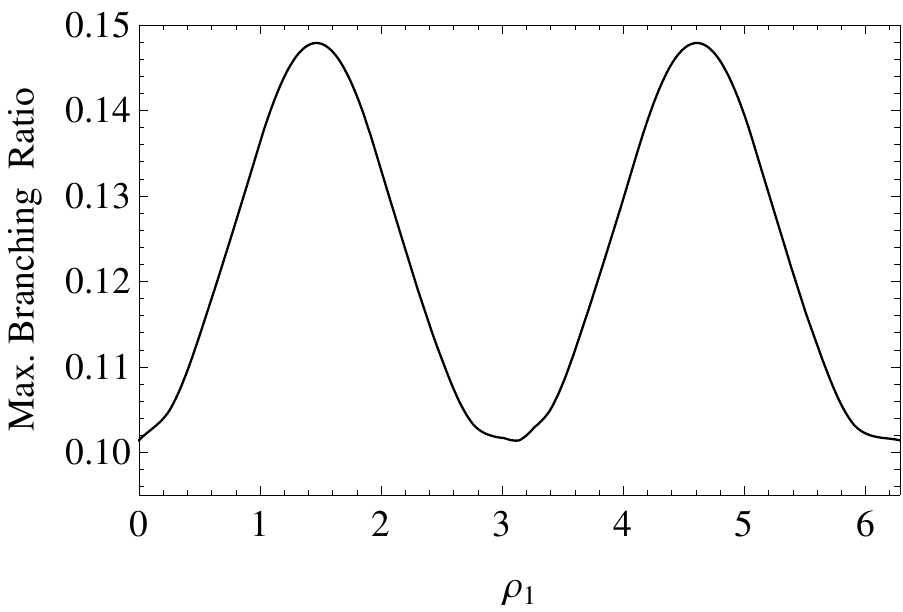}\; \;
\includegraphics[width=0.6\textwidth,angle=0,scale=0.8]{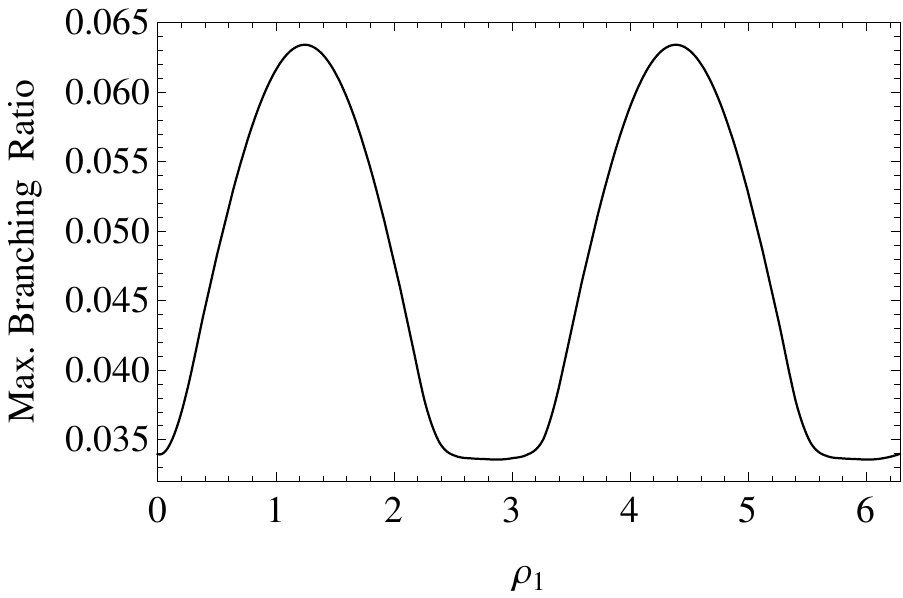}
\end{center}
\caption{In the left (right) panel,  we show the prameter scan results for the maximum allowed branching ratios, $\sum_{i=1}^2{\rm BR}(N_{m}^i N_m^i \rightarrow \mu^\pm \mu^\pm W^\mp W^\mp)$, as a function of a Majorana phase $\rho_1$ for the NH (IH) case. The solid curve denotes the maximum value of the branching ratio obtained after  performing a parameter scan for rest of the free parameters, $X$ and $Y$. From the figure we read the maximum value to be $0.148$ ($0.0634$) for the NH (IH) case. }
\label{scan2}
\end{figure}
%%%%%%%%%%%%%%%
%%%%%%%%%%%%%%%
Let us now consider the alternative U(1)$_X$ model. 
Note that in this model only the first two generation RHNs are involved in the seesaw mechanism (the minimal seesaw \cite{King:1999mb, Frampton:2002qc}). 
In order to make our discussion simple, 
  we assume the degeneracy of the heavy neutrinos in mass such as $m_N= m_N^{1}=m_N^{2}$, and  $m_{N}^{3} \simeq m_{Z^{\prime}}/2$. 
The minimal seesaw scenario predicts one massless light neutrino eigenstate. 
In the NH case, the diagonal mass matrix is given by 
\bea 
  D_{\rm{NH}} ={\rm diag}
  \left(0, \sqrt{\Delta m_{12}^2},
           \sqrt{\Delta m_{12}^2 + \Delta m_{23}^2} \right),  
\label{DNH}
\eea 
while in the IH case 
\bea 
  D_{\rm{IH}} ={\rm diag}
\left( \sqrt{\Delta m_{23}^2 - \Delta m_{12}^2}, 
 \sqrt{\Delta m_{23}^2}, 0 \right).  
\label{DIH}
\eea 
The matrices $\sqrt{D_\nu }$ for the NH and the IH are defined as 
\bea
\sqrt{D_{\rm{NH}}}=
\begin{pmatrix}
0&0\\
 (\Delta m_{12}^2)^{\frac{1}{4}} & 0 \\
0   & (\Delta m_{23}^2+\Delta m_{12}^2)^{\frac{1}{4}} \\
\end{pmatrix},
\sqrt{D_{\rm{IH}}} =
\begin{pmatrix}
    (\Delta m_{23}^2 - \Delta m_{12}^2)^{\frac{1}{4}} & 0 \\
    0 & (\Delta m_{23}^2)^{\frac{1}{4}} \\
    0&0
\end{pmatrix}, 
\eea 
respectively, and $O$ is a general $2\times2$ orthogonal matrix given by
\bea
O=\begin{pmatrix}
\cos(X+i Y)&\sin(X+i  Y)\\
-\sin(X+i Y)&\cos(X+i Y)
\end{pmatrix}
=\begin{pmatrix}
\cosh Y &i \sinh Y\\
-i\sinh Y & \cosh Y
\end{pmatrix}
\begin{pmatrix}
\cos X & \sin  X\\
-\sin X & \cos X
\end{pmatrix}
\eea
where $X$ and $Y$ are real parameters. 
With the inputs of the neutrino oscillation data and $M_N=m_{Z^\prime}/4$ with $m_{Z^\prime}=3$ TeV, 
  we have performed a scan for the free parameters ($X, Y$, and $\rho_1$), 
  and found the maximum values of the branching ratio as  
  $\sum_{i=1}^2{\rm BR}(N_{m}^i N_m^i \rightarrow \mu^\pm \mu^\pm W^\mp W^\mp) \simeq  0.148$ and $0.0634$, 
  for the NH and IH cases, respectively (see Fig.~\ref{scan2}). 
For both the NH and the IH, the maximum values for  the branching ratios are obtained for $\rho_1 \simeq \pi/2$ and $|Y|\gtrsim 2$. 
The result becomes independent of $Y$ for $|Y|\gtrsim 2$.

%%%%%%%%%%%%%%%%%%%%%%%%%
%%%%%%%%%%%%%%%%%%%%%%%%%
\begin{figure}[h]
\begin{center}
\includegraphics[width=0.8\textwidth,angle=0,scale=0.8]{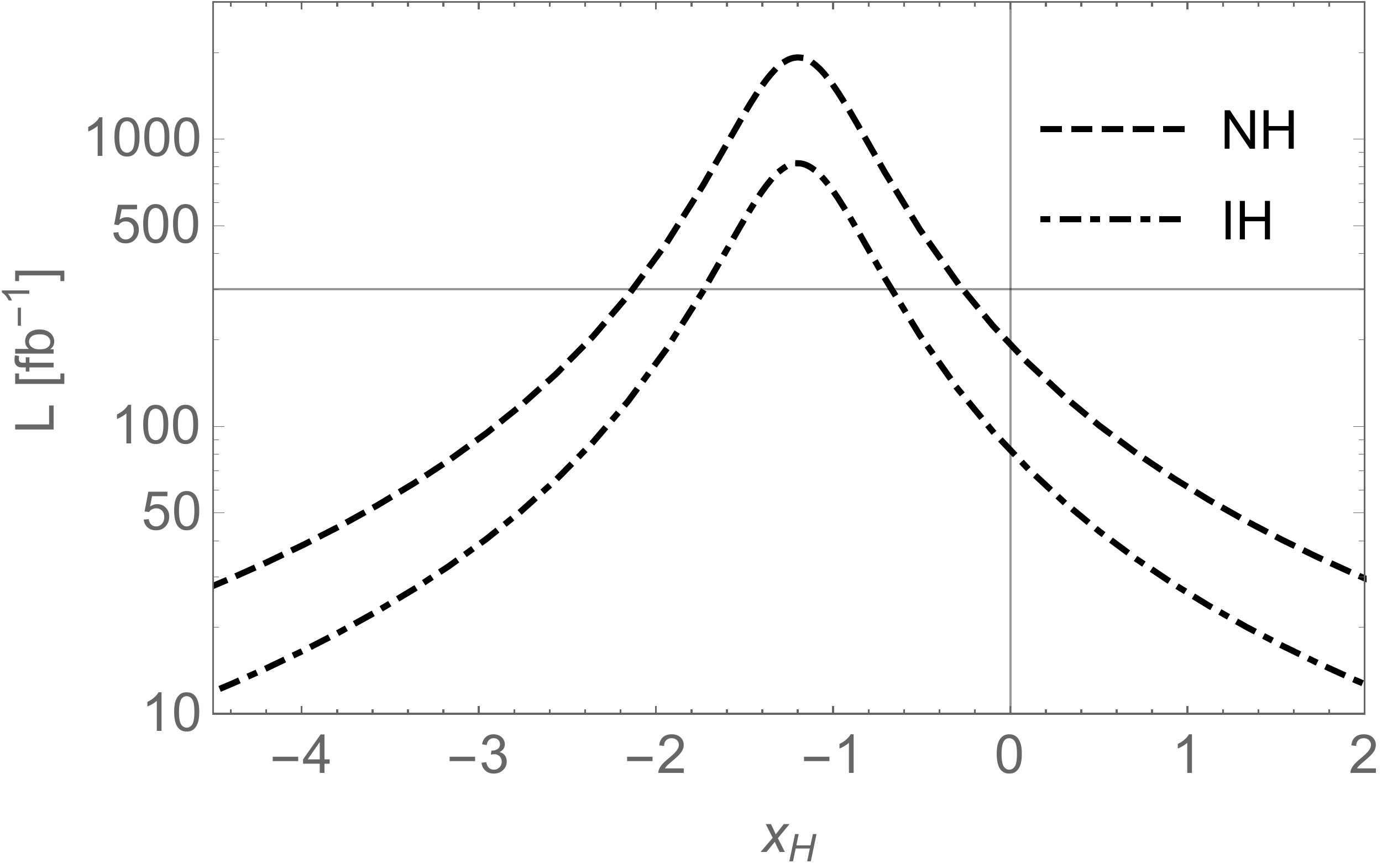}
\end{center}
\caption{The luminosity required to obtain 25 signal events of  the $Z^\prime$ boson as a function of $x_H$ for fixed values of $m_{ Z^{\prime}} = 3$ TeV and  $m_{N^{1,2,3}} = m_{ Z^{\prime}}/4$. 
The dashed (dot-dashed) lines correspond to NH (IH) cases with $\sum_{i=1}^2{\rm BR}(N_{m}^i N_m^i \rightarrow \mu^\pm \mu^\pm W^\mp W^\mp) =0.148$ ($0.0634$). The vertical solid line marks the $B-L$ limit ($x_H = 0$). 
The solid horizontal line corresponds to a luminosity value of 300 ${\rm fb}^{-1}$ required for the discovery of RHNs at the future LHC with a dimuon and a diboson final states. 
}
\label{DiscAlt2}
\end{figure}
%%%%%%%%%%%%%%%
%%%%%%%%%%%%%%%

Using these realistic values for branching ratios to reproduce the neutrino oscillation data, we now re-evaluate the luminosity required for 25 signal events of the $Z^\prime$ boson production. 
For fixed values of $m_{ Z^{\prime}} = 3$ TeV and $m_{N^{1,2}} = m_{ Z^{\prime}}/4$, we show the luminosity as a function of $x_H$  in Fig.~\ref{DiscAlt2}. 
The dashed (dot-dashed) line corresponds to the NH (IH) case. 
Note that with a very large enhancement factor, the alternative U(1)$_X$ model allows us to discover the RHNs at the LHC well before the discovery of the $Z^\prime$ boson. 
For example, for $x_H = -1.2$, using the realistic branching ratios for the RHNs, $\sum_{i=1}^2{\rm BR}(N_{m}^i N_m^i \rightarrow \mu^\pm \mu^\pm W^\mp W^\mp) =0.148$ and  $0.0634$, the required luminosity is found to be ${{\mathcal L}}({\rm fb}^{-1})  \simeq 1923$ and $824$ for the NH and the IH cases, respectively. 
For the $B-L$ limit ($x_H = 0$) case, we previously obtained ${{\mathcal L}}({\rm fb}^{-1}) \simeq 325$, for $\text{BR}(N\to W \mu) \simeq 0.5$. 
Using the realistic branching ratios for the RHNs, the corresponding luminosities are reduced to ${{\mathcal L}}({\rm fb}^{-1})  \simeq 192$ and $82$ for the NH and the IH cases, respectively. 
Accordingly, the allowed range of $x_H$ values for the NH (IH) case is reduced to be  $-4.1 \leq x_H \leq 1.7$ ($-3.1 \leq x_H \leq 0.7$).  
The solid horizontal line corresponds to a luminosity value of 300 ${\rm fb}^{-1}$ required for the discovery of RHNs at the future LHC with a dimuon and a diboson final states. 
Hence for the NH (IH), Fig.~\ref{DiscAlt2} indicates that the RHNs will be discovered before the $Z^\prime$ boson for $-2.1 \lesssim x_H\lesssim 0$ ($-1.7 \lesssim x_H\lesssim -0.7$).

%%%%%%%%%%%%%%
\section{Conclusions}
\label{sec:5}
%%%%%%%%%%%%%%
We have investigated a prospect of discovering the RHNs in type-I seesaw at the LHC, 
   which are pair produced from the decay of a resonantly produced $Z^\prime$ boson. 
Recent simulation studies show that the discovery of the RHNs via $Z^{\prime}\to NN$ 
   is promising at the future LHC with, for example, a 300 fb$^{-1}$ luminosity. 
However, the production cross section of $Z^{\prime}$ boson into dilepton final states 
  ($pp \to Z^{\prime}\to \ell^{+}\ell^{-} $, where $\ell^\pm=e^\pm$ or $\mu^\pm$) 
   is very severely constrained by the current LHC results. 
Imposing this constraint, we have found that a significant enhancement of the branching ratio 
  ${\rm BR}(Z^{\prime}\to NN$) over ${\rm BR}(Z^{\prime}\to \ell^{+}\ell^{-}$) is crucial for the future discovery of RHNs.  
For the minimal gauged U(1)$_X$ extension of the SM 
  with the conventional and the alternative charge assignments, 
  we have found that a significant enhancement, ${\rm BR}(Z^{\prime}\to NN)/{\rm BR}(Z^{\prime}\to \ell^{+}\ell^{-}) \simeq 3.25 $ 
  and 52 (per generation), respectively, can be achieved for $x_H = -1.2$, with $m_{Z^\prime} = 3$ TeV, and $m_N = m_{Z^\prime}/ 4$. 
This is in sharp contrast with the ratio, ${\rm BR}(Z^{\prime}\to NN)/{\rm BR}(Z^{\prime}\to \ell^{+}\ell^{-}) \simeq 0.5$, 
   in the minimal $B-L$ model which is commonly used in the simulation studies. 
The branching ratio of $\text{BR}(N\to W \mu) = 0.5$ is commonly assumed in the simulation studies. 
However, this branching ratio is not consistent with the neutrino oscillation data. 
Employing the general parameterization of the neutrino Dirac mass matrix to reproduce the neutrino oscillation data, we have performed a parameter scan to evaluate the maximal value for $\text{BR}(N\to W \mu)$. 
With the maximum enhancement factors and the maximum branching ratio, 
we have concluded for the minimal U(1)$_X$ model that a 5$\sigma$ discovery of RHNs in the future according to the simulation studies implies that the $Z^\prime$ boson must be discovered before the RHNs. 
In the alternative U(1)$_X$ model, we have obtained further enhancement of the signal cross section than the conventional case, and found a possibility of discovering the RHNs even before the $Z^\prime$ boson at the future LHC experiment.

%%%%%%%%%%%%%%%%%%%%%%%%%%%%%%%%%%%%%%%%%%%%%%%%%%%%%%%%%%%%%%
%\bigskip
\acknowledgments
The work of N.O. is supported in part by the United States Department of Energy (No.~DE-SC0013680).

%%%%%%%%%%%%%%%%%%%%%%%%%%%%%%%%%%%%%%%%%%%%%%%%%%%%%%%%%%

 \end{document}